\documentclass[pdflatex,sn-nature]{sn-jnl}
\usepackage{graphicx} % Required for inserting images
\usepackage{dcolumn}% Align table columns on decimal point
\usepackage{bm}% bold math
\usepackage{xfrac}
\usepackage{mathrsfs}
\usepackage{xcolor}
\usepackage{array}
\usepackage{booktabs}
\usepackage{hyperref}

\usepackage{multirow}%
\usepackage{amsmath,amssymb,amsfonts}%
\usepackage{amsthm}%
\usepackage{mathrsfs}%
\usepackage[title]{appendix}%
\usepackage{textcomp}%
\usepackage{manyfoot}%
\usepackage{algorithm}%
\usepackage{algorithmicx}%
\usepackage{algpseudocode}%
\usepackage{listings}%
\usepackage{orcidlink}
\usepackage{pdfpages}

\usepackage{comment}

\begin{document}
\title[High-throughput computation of electric polarization in solids via Berry flux diagonalization]{High-throughput computation of electric polarization in solids via Berry flux diagonalization}

\author[1,2]{\fnm{Abigail N.} \sur{Poteshman}\orcidlink{0000-0002-4873-4826}}

\author[3, 4, 5, 6]{\fnm{Francesco} \sur{Ricci}\orcidlink{0000-0002-2677-7227}}

\author*[3, 4, 7]{\fnm{Jeffrey B.} \sur{Neaton}\orcidlink{0000-0001-7585-6135}}\email{jbneaton@lbl.gov}

\affil[1]{\orgdiv{Committee on Computational and Applied Mathematics}, \orgname{University of Chicago}, \orgaddress{\city{Chicago}, \state{IL}, \postcode{60637}, \country{USA}}}

\affil[2]{\orgdiv{Materials Science Division}, \orgname{Argonne National Laboratory}, \orgaddress{\city{Lemont}, \state{IL}, \postcode{60439} \country{USA}}}

\affil[3]{\orgdiv{Materials Sciences Division}, \orgname{Lawrence Berkeley National Laboratory}, \orgaddress{\city{Berkeley}, \state{CA}, \postcode{94720} \country{USA}}}

\affil[4]{\orgdiv{Department of Physics}, \orgname{University of California}, \orgaddress{\city{Berkeley}, \state{CA}, \postcode{94720} \country{USA}}}

\affil[5]{\orgdiv{ Institute of Condensed Matter and Nanosciences (IMCN)}, \orgname{Université catholique de Louvain (UCLouvain)}, \orgaddress{ \city{Louvain-la-Neuve}, \country{Belgium}}}

\affil[6]{\orgdiv{Matgenix SRL}, \orgname{A6K Advanced Engineering Centre}, \orgaddress{\city{Charleroi}, \postcode{6000}, \country{Belgium}}}

\affil[7]{\orgdiv{Kavli Energy NanoSciences Institute at Berkeley}, \orgaddress{\city{Berkeley}, \state{CA}, \postcode{94720} \country{USA}}}

\abstract{Electric polarization in the absence of an externally applied electric field is a key property of polar materials, but the standard interpolation-based \textit{ab initio} approach to compute polarization differences within the modern theory of polarization presents challenges for automated high-throughput calculations. Berry flux diagonalization [J. Bonini \textit{et al.}, Phys. Rev. B \textbf{102}, 045141 (2020)] has been proposed as an efficient and reliable alternative, though it has yet to be widely deployed. Here, we assess Berry flux diagonalization using \textit{ab initio} calculations of a large set of materials, introducing and validating heuristics that ensure branch alignment with a minimal number of intermediate interpolated structures. Our automated implementation of Berry flux diagonalization succeeds in cases where prior interpolation-based workflows fail due to band-gap closures or branch ambiguities. Benchmarking with \textit{ab initio} calculations of 176 candidate ferroelectrics, we demonstrate the efficacy of the approach on a broad range of insulating materials and obtain accurate effective polarization values with fewer interpolated structures than prior automated interpolation-based workflows. Our real-space heuristics that can predict gauge stability \textit{a priori} from ionic displacements enable a general automated framework for reliable polarization calculations and efficient high-throughput screening of chemically and structurally diverse polar insulators. These results establish Berry flux diagonalization as a robust and efficient method to compute the effective polarization of solids and to accelerate the data-driven discovery of functional polar materials.}

\maketitle

\section*{Introduction}
\label{sec:intro}

Polar crystals are solids that exhibit a nontrivial electric polarization in the absence of an externally applied electric field. Materials with this property play a central role in a range of electronic, electromechanical, and energy-related applications. For example, ferroelectrics are aclass of polar materials in which the polarization is switchable under an external electric field, with applications in non-volatile memory devices, sensors, and energy storage \cite{scott2007applications, shin2023tunable}. Multiferroics, which feature ferroelectricity coexisting with magnetic order, further expand the scope of applications to include magnetoelectric devices, spintronic devices, and multifunctional sensors \cite{spaldin2019advances}. 

A key step in first principles studies of insulating polar materials is the computation of the effective polarization. Within the modern theory of polarization, the effective polarization is defined as the change in formal polarization along an adiabatic, insulating path connecting a polar material to a nearby nonpolar reference structure \cite{resta2007theory}. The formal polarization of an insulator is computed via the Berry phase of occupied electronic states. The formal polarization is an inherently multi-valued quantity, since it is defined modulo a quantum of polarization set by the unit cell volume and lattice vectors. Although the Berry phase is now a routine output of first-principles electronic structure codes, obtaining the effective polarization requires taking differences between multi-valued formal polarizations and tracking the correct branch of the Berry phase across the chosen path. As a result, automated high-throughput calculations of effective polarizations remain challenging, particularly in cases where the effective polarization is comparable in magnitude to the polarization quantum \cite{neaton2005first}.

Neither the formal nor the effective polarization are quantities that can be measured directly. Experimental studies typically characterize spontaneous polarization, which can be inferred from the integrated switching current that results when an external electric field reverses the polarization of a material. The spontaneous polarization is related to half of the charge transported during a switch between two polar enantiomorphs. The nonpolar reference structure used in the theoretical definition of effective polarization is not intended to represent a physical configuration that the material visits during switching but is instead a practical construction indicating a halfway point between two polar enantiomorphs along a single symmetric path \cite{vanderbilt2018berry}.  Effective polarizations obtained in this way can be compared to experimental spontaneous polarizations, although experimental values may depend on non-bulk factors such as surface terminations and interfaces with electrodes. While several considerations arise in selecting an appropriate nonpolar reference structure that can be connected to a give polar structure along an insulating path, here we focus on the challenge of computing differences in formal polarization along the proper polarization branch in a consistent, efficient, and automatic manner. For discussion of automated strategies for identifying suitable nonpolar reference structures, we refer the reader to Ref. \cite{ricci2024candidate}.

The standard approach to computing effective polarization relies on linear interpolation of atomic coordinates and repeated Berry-phase evaluations to compute formal polarizations along an insulating, adiabatic path \cite{smidt2020automatically,ricci2024candidate}. Although robust for simple systems, this procedure is computationally costly and can be unreliable when automated. The required number of intermediate configurations to resolve polarization branches depends on system-specific parameters, such as the polarization quantum and the unknown value of effective polarization itself. If too few interpolation structures are used, the branch cannot be properly traced, but adding more such structures increases the computational cost. Moreover, intermediate structures may exhibit spurious band gap closures, leading to metallic states for which the Berry phase is ill defined. Born effective charges, obtained either via density functional perturbation theory (DFPT) or finite differences calculations, have previously been used to guide the choice of the correct branch or as a direct proxy to the polarization itself, but they come at greater computational cost and often with lesser accuracy \cite{resta1994macroscopic, king1993theory}. Adaptive interpolation schemes (e.g., the nudged elastic band (NEB) method \cite{munro2019implementation}), though viable, can be even more computationally expensive due to the need for repeated \textit{ab initio} calculations and Berry phase evaluations. 

An alternative approach to computing effective polarization, known as Berry flux diagonalization, was recently proposed by Bonini \textit{et al.} \cite{bonini2020berry} and can circumvent these challenges. Rather than interpolating along a real-space structural path between two structures, Berry flux diagonalization directly computes the differences in formal polarization along the proper branch between two insulators \cite{bonini2020berry}. Due to the underlying periodicity of Bloch wavefunctions in crystalline solids, there is a direct correspondence between branch-hopping in polarization and non-zero winding numbers of Berry phase differences \cite{vanderbilt2018berry}. By constructing a gauge that aligns wavefunctions across $k$-points in the Brillouin zone and structures, formal polarization differences can be computed without explicit interpolations of atomic positions. 

Initial implementations of Berry flux diagonalization \cite{bonini2020berry} were limited to single-direction effective polarization calculations and small test sets, and lacked numerical heuristics to guarantee gauge stability. Consequently, the method, while conceptually elegant, has yet to be broadly adopted or validated for diverse material classes, including systems involving large displacements. In this work, we extend, systematize, and automate an implementation of Berry flux diagonalization to enable robust and efficient effective polarization calculations across a wide range of polar insulators. We generalize the implementation of Berry flux diagonalization to compute polarization automatically along all three directions and support spin-polarized systems. Additionally, we provide a generalizable and extensible workflow that, while currently demonstrated for \textsc{VASP} and \textsc{Quantum ESPRESSO}, can be compatible with any plane-wave based density functional thoery (DFT) code. Crucically, we introduce a physical heuristic relating atomic displacements to wavefunction overlap singular values, providing an \textit{a priori} criterion for when adiabatic gauge alignment is likely to succeed. We benchmark our automated implementation of the Berry flux diagonalization method with the developed heuristics against the standard interpolation-based approach on 176 candidate ferroelectric materials that were previously identified in \cite{ricci2024candidate}, demonstrating that our implementation of Berry flux diagonalization accelerates the calculation of effective polarization while keeping the same accuracy as standard approaches and providing a foundation for large-scale screening and discovery of new polar materials.

\begin{figure}[t]
    \centering
    \includegraphics[width=\textwidth]{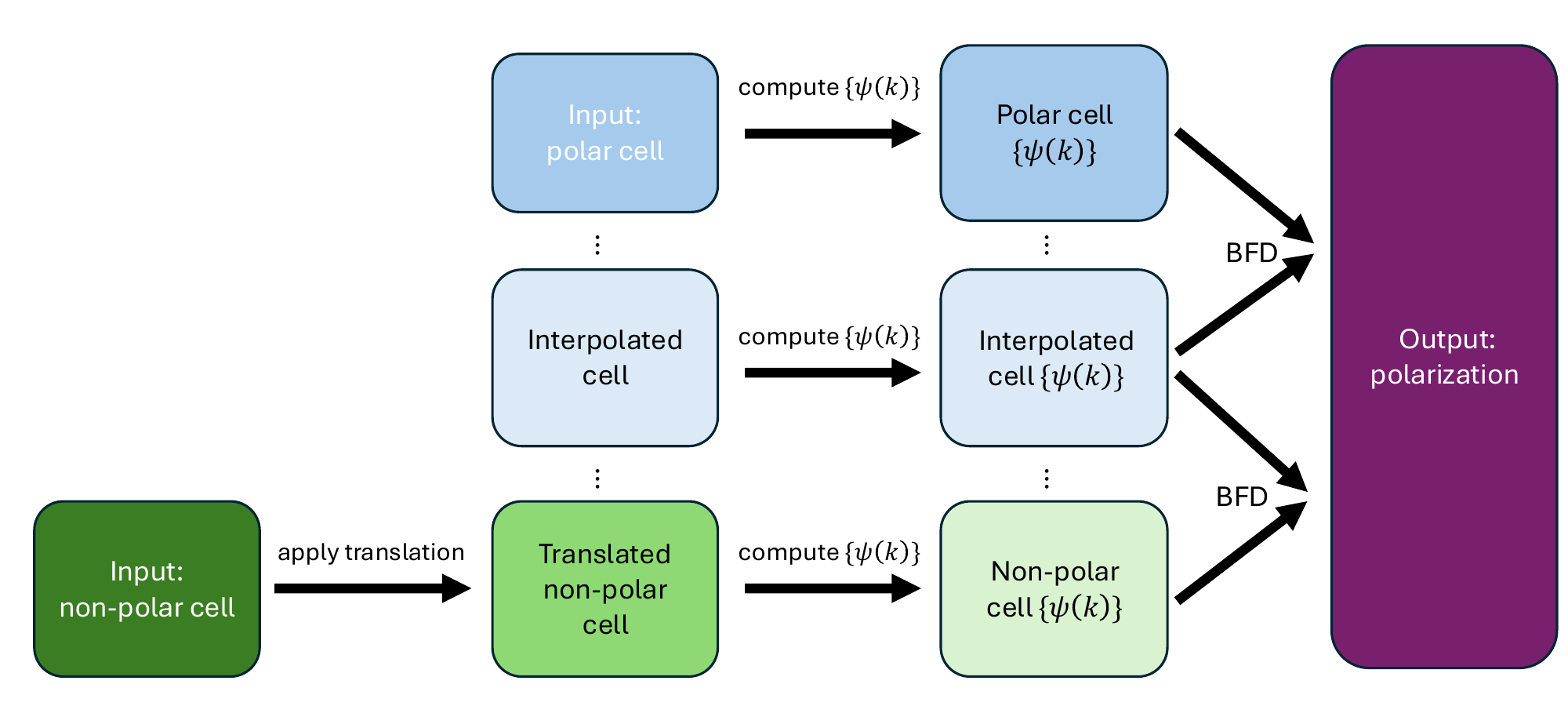}
    \caption{Schematic of our automated workflow to compute effective polarization. First, a rigid translation is applied to the nonpolar reference cell to minimize maximal atomic displacement relative to the polar cell. If the maximal atomic displacement between the polar and translated nonpolar reference cell is larger than 0.3 \AA, intermediate structures are created by linearly interpolating the atomic coordinates. Then, Bloch wavefunctions for a uniform $k$-grid are computed using DFT. The Berry flux diagonalization algorithm is directly applied to wavefunctions corresponding to the polar and nonpolar cells (and any intermediates) to obtain a value for polarization.}
    \label{fig:schematic}
\end{figure}

\section*{Theoretical formalism}
\label{subsec:bfd_method}

We begin provide a brief summary of the standard Berry phase approach to computing effective polarization within the modern theory of polarization, and we provide a review of the Berry flux diagonalization method to compute effective polarization. In our presentation of the standard Berry phase approach method to compute effective polarization and the Berry flux diagonalization formalism, we emphasize the mathematical connections between the two approaches, and we highlight the equations that inform the heuristics we develop for our implementation of an automated workflow using Berry flux diagonalization to compute effective polarization.
For detailed background and pedagogy, we refer the reader to Refs. \cite{king1993theory, resta1994macroscopic, spaldin2012beginner} for the modern theory of polarization and to Ref. \cite{bonini2020berry} for Berry flux diagonalization. 

\subsubsection*{Standard approach for computing polarization using DFT}
\label{subsubsec:standard_approach}

 In the modern theory of polarization, the formal polarization ($\mathbf{P}$) of a crystalline solid, consisting of both ionic and electronic contributions, can be written as
\begin{equation}
   \mathbf{P} = \mathbf{P}_{\text{elec}} + \mathbf{P}_{\text{ion}} \mod \mathbf{P}_\text{quanta}, 
\end{equation} where $\mathbf{P}_\text{quanta}$ is the polarization quantum, an integer multiple of $\sfrac{e\mathbf{R}}{\Omega_{\text{cell}}}$, which is related to the periodicity of the wavefunction and ionic coordinates under translation by a lattice vector. The polarization quanta for a lattice is given by

\begin{equation}
\mathbf{P}_\text{quanta} = \sum_{i \in {a,b,c}} n_i \frac{e\mathbf{R}_i}{\Omega_{\text{cell}}},
\end{equation} where $\Omega_{\text{cell}} $ is the volume of the unit cell, $\mathbf{R}_i$ are primitive lattice vectors, and $n_i \in \mathbb{Z}$. The formal polarization $\mathbf{P} $ of a crystalline solid is therefore a lattice of values and any one value can only be defined modulo a quantum. In practice, only differences in polarization ($\Delta \mathbf{P}$) between related structures are meaningful \cite{resta1994macroscopic}, and as long as the difference in structures are small enough one can always select a common reference quantum, or branch, that yields a single well-defined $\Delta \mathbf{P}$. We will return to this issue later.

In practice, the ionic contribution, $\Delta \mathbf{P}_{\text{ion}}$, can be computed from the difference in atomic coordinates as 
\begin{equation}
\label{eq:ionic_contrib}
    \Delta \mathbf{P}_{\text{ion}} = \frac{e}{\Omega_{\text{cell}}}\sum_{i} Z_i \Delta \mathbf{r}_i,
\end{equation} where for each atom indexed by $i$, $Z_i$ is the ionic charge, defined as the total atomic number of the atom minus the number of its electrons explicitly treated as valence by the pseudopotentials in the calculation, and $\Delta \mathbf{r}_i$ is the displacement of the atom between the polar and nonpolar reference structures.

The electronic contribution to the formal polarization for a given atomic arrangement is given by the Berry phase $\gamma_n$ of occupied Bloch states, namely
 
 \begin{equation}
 \mathbf{P}_{\text{elec}} = \frac{e}{2\pi} \sum_n \gamma_n, 
 \end{equation} where $n$ is over occupied states, and the electronic contribution to effective polarization is computed as the difference between the formal polarization of the polar (pol) and nonpolar (np) structures:

 \begin{equation} 
 \label{eq:pol_elec_diff}
 \Delta \mathbf{P}_{\text{elec}} = \frac{e}{2 \pi}\mathbf{\Phi} = \mathbf{P}_{\text{elec}}^{\text{pol}} - \mathbf{P}_{\text{elec}}^{\text{np}}. 
 \end{equation} The Berry phase $ \gamma_n $ is typically computed as the integral of the Berry connection $\mathbf{A}_n(\mathbf{k})$ along a closed path $ \mathcal{C} $ in reciprocal space, that is

\begin{equation}
\label{eq:gamma_path_int}
\gamma_n = \oint_{\mathcal{C}} \mathbf{A}_n(\mathbf{k}) \cdot d\mathbf{k},
\end{equation} where

\begin{equation}
\mathbf{A}_n(\mathbf{k}) = \text{Im} \langle u_{n \mathbf{k}} | \nabla_{\mathbf{k}} | u_{n \mathbf{k}} \rangle.
\end{equation} Here, $ | u_{n \mathbf{k}} \rangle $ is the cell-periodic part of a Bloch wavefunction and $ \nabla_{\mathbf{k}} $ denotes the gradient with respect to the crystal momentum $ \mathbf{k} $. For a discrete set of $ \mathbf{k} $-points in the Brillouin zone, Eq. \ref{eq:gamma_path_int} becomes

\begin{equation}
\label{eq:overlp_standard}
    \gamma_n = \text{Im} \ln \prod_{m=1}^{N-1} \langle u_{n \mathbf{k_{m+1}}} | u_{n \mathbf{k_m}} \rangle.
\end{equation} The multi-valued nature of formal polarization originates with $\gamma_n$ in Eq. \ref{eq:overlp_standard} being itself multi-valued, due to the fact that the imaginary part of the logarithm of a phase is defined only modulo $2 \pi$. This multi-valuedness is a consequence of the periodicity of the wavefunction under translations by a lattice vector.

The effective polarization is finally determined by subtracting the Berry phase of the nonpolar structure from that of the polar one (Eq. \ref{eq:pol_elec_diff}). However, due to the multi-valued nature of the Berry phase, directly taking this difference can be ambiguous. To ensure the subtraction occurs on a consistent branch of the Berry phase, where both values of the formal polarization in the difference refer to the same quantum, it is common practice to compute the formal polarizations for intermediate structures by interpolating atomic positions between the polar and nonpolar configurations \cite{neaton2005first, smidt2020automatically}. Alternatively, another practice is to estimate the effective polarization via
\begin{equation}
    \Delta \mathbf{P} \cong \frac{1}{\Omega} \sum_i Z_i^* \Delta \mathbf{u}_i,
\end{equation} where the sum runs over all atoms in the unit cell, $i$ indexes an atom, and $\{ Z_i^*\}$ are the Born effective charges of, e.g., the nonpolar or polar phase. This estimate, which is approximate but becomes exact in the limit of small $\Delta \mathbf{u}$, can guide the manual choice of a consistent branch for both phases \cite{resta1994macroscopic, king1993theory}.

\subsubsection*{Berry flux diagonalization}
\label{subsubsec:bfd}

\begin{figure}[t]
    \centering
    \includegraphics[width=0.75\textwidth]{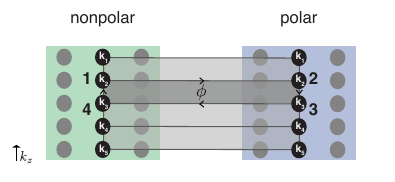}
    \caption{Schematic of a Berry phase ``plaquette'' used in the BFD calculation of Berry phase differences $\phi_n^p$ along the $z$-direction. Each plaquette is defined by a closed $k$-point loop $\mathcal{P}$ involving adjacent $k$-points from both the polar and nonpolar structures. To compute $\phi$ for a single plaquette, we compute overlap matrices between Bloch wavefunctions along the path $\mathcal{P} = (1) \, \mathbf{k}_2^{\text{np}} \rightarrow (2) \, \mathbf{k}_2^{\text{pol}} \rightarrow (3) \, \mathbf{k}_3^{\text{pol}} \rightarrow (4) \, \mathbf{k}_3^{\text{np}} \rightarrow (1) \, \mathbf{k}_2^{\text{np}}$. That is, we compute overlap matrix $M^{\langle 1, 2 \rangle}$ corresponding to the overlap between Bloch wavefunctions at $k$-points labeled  (1) $\mathbf{k}_2^{\text{np}}$ and (2) $\mathbf{k}_2^{\text{pol}}$, and then we compute overlap matrix $M^{\langle 2, 3 \rangle}$ corresponding to the overlap between Bloch wavefunctions at $k$-points labeled (2) $\mathbf{k}_2^{\text{pol}}$ and (3) $\mathbf{k}_3^{\text{pol}}$ in the polar cell, and likewise for $M^{\langle 3, 4 \rangle}$ and $M^{\langle 4, 1 \rangle}$. The Berry phase difference $\phi$ corresponding to this plaquette is computed via Eqs. \ref{eq:overlap_matrix}-\ref{eq:berry_phase_eigs}, and the Berry phase differences corresponding to all plaquettes defined by the $k$-points are summed to obtain the total Berry phase difference $\mathbf{\Phi}$.}  
    
    \label{fig:plaquettes}
\end{figure}

Berry flux diagonalization computes the Berry phase differences (denoted $\phi_n^p$) directly from the Bloch wavefunctions of the two structures (see Fig. \ref{fig:schematic}), as opposed to the standard procedure of computing the Berry phases for each structure independently and then taking their difference. To compute the electronic contribution to effective polarization using Berry flux diagonalization \cite{bonini2020berry}, the total phase difference $\mathbf{\Phi}$ (as in Eq.~\ref{eq:pol_elec_diff}) is written as a sum over Berry phase ``plaquettes'':
\begin{equation}
    \mathbf{\Phi} = \sum_p \sum_n \phi_n^p,
    \label{eq:sum_plaquettes}
\end{equation}
where $n$ indexes the occupied bands and $p$ labels the plaquettes.

Each plaquette is defined by a closed loop in $k$-space involving two adjacent $k$-points in both the polar (pol) and nonpolar (np) structures:
\begin{equation}
\label{eq:path}
    \mathcal{P}(\text{np, pol)} = (1) \, \mathbf{k}_{j}^{\text{np}} 
    \rightarrow (2) \, \mathbf{k}_{j}^{\text{pol}} 
    \rightarrow (3) \, \mathbf{k}_{j+1}^{\text{pol}} 
    \rightarrow (4) \, \mathbf{k}_{j+1}^{\text{np}} 
    \rightarrow (1) \, \mathbf{k}_{j}^{\text{np}},
\end{equation}
where $\mathbf{k}_{i}^{\text{pol}}$ and $\mathbf{k}_{i}^{\text{np}}$ denote $k$-points in the polar and nonpolar structures, respectively (see Fig. \ref{fig:plaquettes}). The points $\mathbf{k}_j$ and $\mathbf{k}_{j+1}$ are adjacent along the string of $k$-points in the direction of the polarization component being computed. Similar to the standard Berry phase approach, this process is repeated along strings of $k$-points, and the effective polarization is obtained by averaging over strings oriented along the complementary reciprocal lattice directions.

To compute the Berry phase $\phi_n^p$ for a given band $n$ and plaquette $p$, we first define overlap matrices between Bloch wavefunctions at adjacent vertices of the plaquette. The Bloch states $|u_{n}(\mathbf{k}_a)\rangle$ are labeled by band index $n$ and $k$-point $\mathbf{k}_a$, where $a$ is a $k$-point along the path of $\mathcal{P}$ (Eq. \ref{eq:path}). The overlap matrix between two adjacent points $\mathbf{k}_a$ and $\mathbf{k}_b$ on the path $\mathcal{P}$ is
\begin{equation}
\label{eq:overlap_matrix}
    M^{\langle a,b \rangle}_{mn} = \langle u_m(\mathbf{k}_a) | u_n(\mathbf{k}_b) \rangle,
\end{equation}
where $m$ and $n$ are band indices.

Each overlap matrix $M^{\langle a,b \rangle}$ is then decomposed using a singular value decomposition (SVD) as:
\begin{equation}
\label{eq:svd}
    M^{\langle a,b \rangle} = V^{\langle a,b \rangle} \Sigma^{\langle a,b \rangle} W^{\langle a,b \rangle \dagger},
\end{equation}
where $V^{\langle a,b \rangle}$ and $W^{\langle a,b \rangle}$ are unitary matrices, and $\Sigma^{\langle a,b \rangle}$ is a diagonal matrix of singular values. We then compute the unitary approximant of each overlap matrix from its SVD by discarding the singular values, expressed as
\begin{equation}
    \mathcal{M}^{\langle a,b \rangle} = V^{\langle a,b \rangle} W^{\langle a,b \rangle \dagger}.
\end{equation}

The total unitary matrix $U_{\mathcal{P}}$ that describes evolution around plaquette $p$ is constructed as the ordered product of unitary approximants along the four edges of the plaquette path $\mathcal{P}$, namely
\begin{equation}
\label{eq:unitary_approximant}
    U_{\mathcal{P}} = \mathcal{M}^{\langle 1,2 \rangle} \mathcal{M}^{\langle 2,3 \rangle} \mathcal{M}^{\langle 3,4 \rangle} \mathcal{M}^{\langle 4,1 \rangle}.
\end{equation}
The eigenvalues of this unitary matrix encode the geometric phase acquired by each band, and can be expressed as
\begin{equation}
\label{eq:eig_decomp}
    \text{eig}(U_{\mathcal{P}}) = \left\{ e^{i \phi_n^p} \right\};
\end{equation}
the phases themselves are extracted as
\begin{equation}
\label{eq:berry_phase_eigs}
    \phi_n^p = \text{Im} \ln \text{eig}(U_{\mathcal{P}}).
\end{equation}

Importantly, this approach requires both intra-structure overlap matrices (e.g., $\mathcal{M}^{\langle 2,3 \rangle}$ and $\mathcal{M}^{\langle 4,1 \rangle}$) and inter-structure ones (e.g., $\mathcal{M}^{\langle 1,2 \rangle}$ and $\mathcal{M}^{\langle 3,4 \rangle}$). The inter-structure overlaps are not present in standard Berry phase calculations, but they are essential here to evaluate the differences in formal polarizations directly and consistently across branches.

Berry flux diagonalization can circumvent branch-hopping so long as the singular values of the overlap matrices (Eq.~\ref{eq:svd}) is greater than zero and each $\phi_i^p \ll \pi$. The singular value requirement ensures that the phase in Eq.~\ref{eq:berry_phase_eigs} is not extracted from a complex number whose magnitude is zero. The eigenvalue requirement is equivalent to ensuring that the phase in the eigenvalues $e^{i\phi_n^p}$ do not wrap around the unit circle (i.e., they have a non-zero winding number). In the context of the modern theory of polarization, which is formulated within the adiabatic approximation, this eigenvalue condition corresponds to ensuring that the wavefunctions evolve smoothly along an adiabatic path in parameter space. That is, wavefunctions at adjacent $k$-points or corresponding $k$-points across the polar and nonpolar structure, must start sufficiently close to each other so that they are adiabatically connected. To ensure that both of these conditions are satisfied in practice, we develop a real-space heuristic based on the difference in ionic positions, $\Delta \mathbf{r}_i$, from Eq. \ref{eq:ionic_contrib}, in polar and nonpolar structures. In the next section, we connect the real-space displacement of ions, the singular values of the overlap matrices (Eq.\ref{eq:svd}), and the eigenvalues of the plaquette evolution matrix (Eq.\ref{eq:eig_decomp}). This analysis enables us to formulate a criterion to ensure the Berry phase contributions are numerically stable, branch-consistent, and avoid the breakdown of adiabaticity.

\subsection*{Heuristic for sufficient wavefunction overlap and gauge consistency}
\label{subsec:heuristic}
To compute the difference in Berry phase along the correct branch of polarization using Berry flux diagonalization, singular values of the overlap matrices (Eq. \ref{eq:svd}) must be greater than zero and the eigenvalues of the Berry flux plaquettes (Eq. \ref{eq:berry_phase_eigs}) must be less than $\pi$. When the singular values of the overlap matrices are zero, we are in a numerically unstable regime in which we are extracting phases from complex numbers with zero magnitude. When eigenvalues of the Berry flux plaquettes reach or exceed $\pi$, the Berry flux diagonalization formalism fails and is equivalent to jumping between polarization branches. Bonini et al \cite{bonini2020berry} suggested that maintaining a minimum singular value $> 0.15$ in the overlap matrices (Eq. \ref{eq:svd}) is sufficient to preserve proper branch alignment. Here, we propose a complementary real-space heuristic that requires only atomic coordinates: a maximal atomic displacement of $\approx 0.3$ \AA  \ between polar and nonpolar structures reliably yields minimum singular values $> 0.15$ and maximum eigenvalues $\ll \pi $ and thus ensures proper numerical stability and gauge continuity. We systematically test this heuristic  using a system with artificially displaced atoms, translations of atomic coordinates for a set of standard ferroelectrics, and a deterministic number of interpolations to reduce atomic displacements in cases where atomic displacements are large ($> 0.3$ \AA). The results in this section are all presented for calculations using \textsc{VASP}, but the same calculations are also performed in \textsc{Quantum ESPRESSO}, and we present those results in Sec. V of the Supplementary Information.

\begin{figure}[t!]
    \centering
    \includegraphics[width=0.8\textwidth]{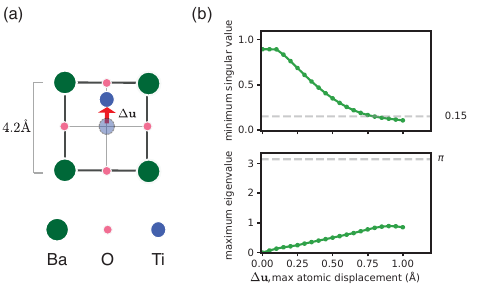}
    \caption{(a) Schematic of cubic perovskite BaTiO$_3$ test system with fixed Ba and O atoms and a Ti atom that is artificially displaced along the [001] direction (red arrow). The displacement of the Ti atom, $\Delta \mathbf{u}$, is the difference between its artificial position and its centrosymmetric position. (b) Minimal singular values of the overlap matrices $M^{\langle i, i+1 \rangle}$ and maximal eigenvalues of the unitary evolution matrices $U_{P}$ corresponding to the effective polarization calculated using Berry flux diagonalization for the BaTiO$_3$ with the artificially displaced Ti atom.}
    \label{fig:toy-batio3}
\end{figure}

\subsubsection*{Artificial BaTiO$_3$ test system}
\label{subsubsec:toy_batio3}
To illustrate the relationship between nuclear displacement, the minimum singular values of Eq. \ref{eq:svd}, and maximum eigenvalues of Eq. \ref{eq:eig_decomp}, we construct a cubic perovskite BaTiO$_3$ system in which we initialize two identical nonpolar configurations fixed with Ba and O atoms. In one of the nonpolar configurations, we artificially displace only the Ti atom from its centrosymmetric position in the unit cell up to 1 \AA \  along the [001] direction (see Fig. \ref{fig:toy-batio3}a). We compute overlap matrices between the nonpolar cell and the cell with the displaced Ti atom using Berry flux diagonalization, and we report the minimal singular values over all overlap matrices (Eq. \ref{eq:svd}) and maximal loop eigenvalues over all unitary evolution matrices (Eq. \ref{eq:eig_decomp}) in Fig. \ref{fig:toy-batio3}b.

We observe a monotonic decrease in the minimal singular values with increasing displacement of the Ti atom relative to its centrosymmetric position in the cubic perovskite structure. The threshold singular value of 0.15 corresponds to a Ti displacement of approximately 0.75 Å. 
% For structures with singular values below this threshold, the maximal eigenvalues of the Berry flux loops approach or reach $\pi$, signaling a failure to enforce the necessary gauge alignment for Berry phase calculations. This study of an artificial BTO system demonstrates the utility of our heuristic in identifying cases where the Berry flux diagonalization method may fail due to insufficient wavefunction overlap. 
This set of calculations establishes that a quantitative correspondence between ionic displacement in real space, the minimum singular values of the overlap matrices, and the behavior of the maximal eigenvalues of the unitary evolution matrices. We also see that the $0.3$ \AA \ heuristic for maximal atomic displacement and the 0.15 heuristic for the minimal singular value is well below the corresponding threshold for which the maximal eigenvalue approaches $\pi$.

\begin{figure}[h!]
    \centering
    \includegraphics[width=0.6\textwidth]{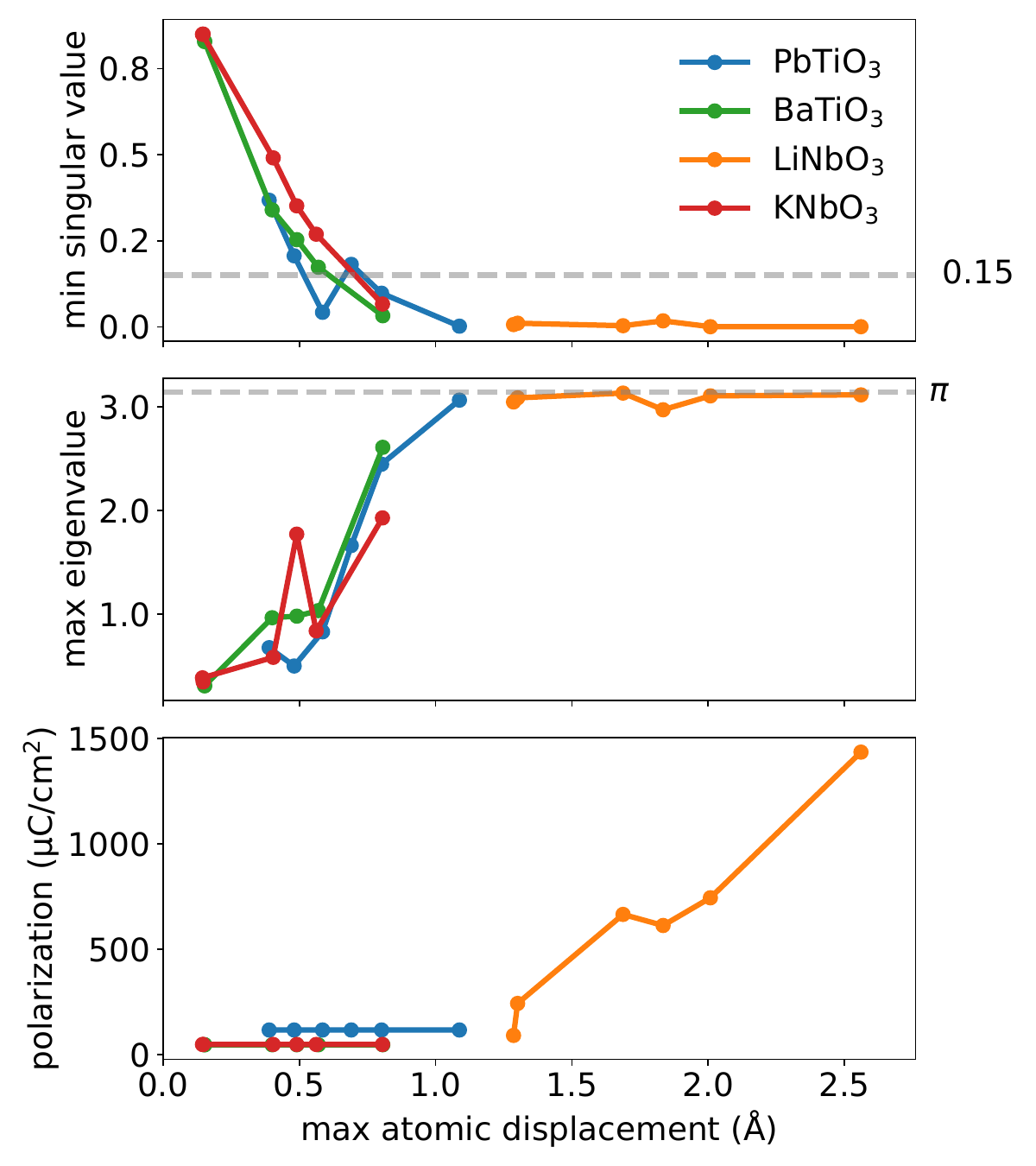}
    \caption{Minimal singular values (top), maximal eigenvalues (middle), and effective polarization (bottom) computed from overlap matrices for standard ferroelectrics under different translations of the nonpolar structure relative to the polar structure. For BaTiO$_3$, KNbO$_3$, and PbTiO$_3$, the translations that minimize the maximal atomic displacement ensure sufficient overlap between wavefunctions at each $k$-point, with minimal singular values remaining above the threshold of 0.15 and maximal eigenvalues below $\pi$. These conditions guarantee proper branch alignment and accurate effective polarization calculations. For LiNbO$_3$, even the translation minimizing maximal atomic displacement results in insufficient overlap, with singular values approaching zero and eigenvalues near $\pi$, leading to improper resolution of polarization branches and inaccurate values of effective polarization. This can be addressed by adding interpolated structures, see main text. Note: polarization values for BaTiO$_3$ (45.8 $\mu C / \text{cm}^2$) are almost identical to those of KNbO$_3$ (47.3 $\mu C / \text{cm}^2$), so the green curve is obscured by the red one in the effective polarization plot. }
    \label{fig:svd-eigs}
\end{figure}

\subsubsection*{Translations operations to minimize maximal atomic displacements}
\label{subsubsec:translations}

To ensure sufficient overlap between wavefunctions of the polar and nonpolar structures in the Berry flux diagonalization approach, we introduce a pre-processing step where the translation between the structures is optimized to minimize the maximal atomic displacement. Specifically, we search for a translation vector $T = (T_a, T_b, T_c) $ such that the maximal atomic displacement along any single direction, $ \max_i ||\Delta \mathbf{r}_i||_{\infty} $, is minimized over all atoms $ i $ in the primitive cell. This pre-processing step, performed before any DFT calculation, preserves the cell parameters and interatomic spacings but alters the origin of the nonpolar structure to reduce the distances between corresponding atoms in the polar and nonpolar configurations, reducing wavefunction mismatch.

Figure \ref{fig:svd-eigs} illustrates the impact of this pre-processing translation step on the minimal singular values and maximal eigenvalues of the overlap matrices for BaTiO$_3$, KNbO$_3$, PbTiO$_3$, and LiNbO$_3$ computed directly from the wavefunctions corresponding to the polar and nonpolar structures (see Sec. I of the SI for the cell parameters of the polar and nonpolar structures). For BaTiO$_3$, KNbO$_3$, and PbTiO$_3$, optimal translations result in sufficiently small maximal atomic displacements, leading to high minimal singular values (above 0.15) and low maximal eigenvalues (below $\pi$), ensuring proper gauge alignment and accurate effective polarization calculations (see Fig. \ref{fig:svd-eigs}a). Furthermore, we observe that increases in the maximal atomic displacements result in decreases in the minimal singular values and increases in maximal eigenvalues. In contrast, for LiNbO$_3$, even the translation that minimizes maximal atomic displacement results in values of minimal singular values of the overlap matrices approaching zero and maximal eigenvalues of the unitary evolution matrix near $ \pi $ (see Fig. \ref{fig:svd-eigs}b), resulting in a polarization of $\sim90  \ \mu C / \text{cm}^2$, which is higher than the reference value of 84.5 $\mu C / \text{cm}^2$ computed from the standard interpolation-based approach. Therefore, this case clearly demonstrates a need to use interpolated structures to further reduce maximal atomic displacement, restore gauge alignment, and recover accurate effective polarizations when a single translation does not result in sufficiently small ionic displacements. We find that adding intermediate interpolated structures for LiNbO$_3$ indeed resolves these numerical instability issues and results in a polarization value (83.4 $\mu C / \text{cm}^2$) in close agreement with the reference value (84.5 $\mu C / \text{cm}^2$).

 \subsubsection*{Interpolated structures}
 \label{subsubsec:interpolations}
For materials where the minimum atomic displacement between polar and nonpolar structures remains relatively large ($>0.3$ \AA), we can introduce intermediate (interpolated) structures along the path connecting the two states, along the lines of the standard Berry phase approach. By decreasing the maximal atomic displacement between adjacent structures below the heuristic value of 0.3 \AA, we can increase the minimum singular values above 0.15 and obtain maximum eigenvalues that are below $\pi$. In cases where a single translation does not sufficiently reduce the maximum atomic displacement, linearly interpolating atoms between the polar and nonpolar structures and computing the Berry phase along this path can minimize structural differences and ensure gauge alignment for Berry flux diagonalization.

We modify the Berry flux diagonalization formalism \cite{bonini2020berry} to compute the electronic contribution to polarization using intermediate, interpolated structures by considering the set of $n-1$ structures \{nonpolar ($i_0$), interpolation$_1$ ($i_1$), interpolation$_2$ ($i_2$),\dots, interpolation$_i$ ($i_i$), \dots, polar ($i_n$)\}. We can re-write the sum in Eq. \ref{eq:sum_plaquettes} as a sum over the plaquettes formed by adjacent sets of interpolated structures, namely 

\begin{equation}
    \mathbf{\Phi} = \sum_{i=0}^{n-1} \sum_{p(i, i+1)} \sum_n \phi_n^{p(i, i+1)},
    \label{eq:sum_plaquettes_interps}
\end{equation} where the plaquettes $p(i, i+1)$ are defined by the path:
\begin{equation}
\label{eq:path_interps}
    \mathcal{P}(\text{i, i+1)} = (1) \, \mathbf{k}_{j}^{i_i} 
    \rightarrow (2) \, \mathbf{k}_{j}^{i_{i+1}} 
    \rightarrow (3) \, \mathbf{k}_{j+1}^{i_{i+1}} 
    \rightarrow (4) \, \mathbf{k}_{j+1}^{i_i} 
    \rightarrow (1) \, \mathbf{k}_{j}^{i_i},
\end{equation} where $i_i$ are the indices over the interpolated structures. In this modified definition of Eq. \ref{eq:path}, $\mathbf{k}_j$ and $\mathbf{k}_{j+1}$ are still adjacent $k$-points along a string in the direction of the polarization component being computed, and $n$ is still the index of the occupied bands. 

\begin{figure}[t!]
    \centering
    \includegraphics[width=0.6\textwidth]{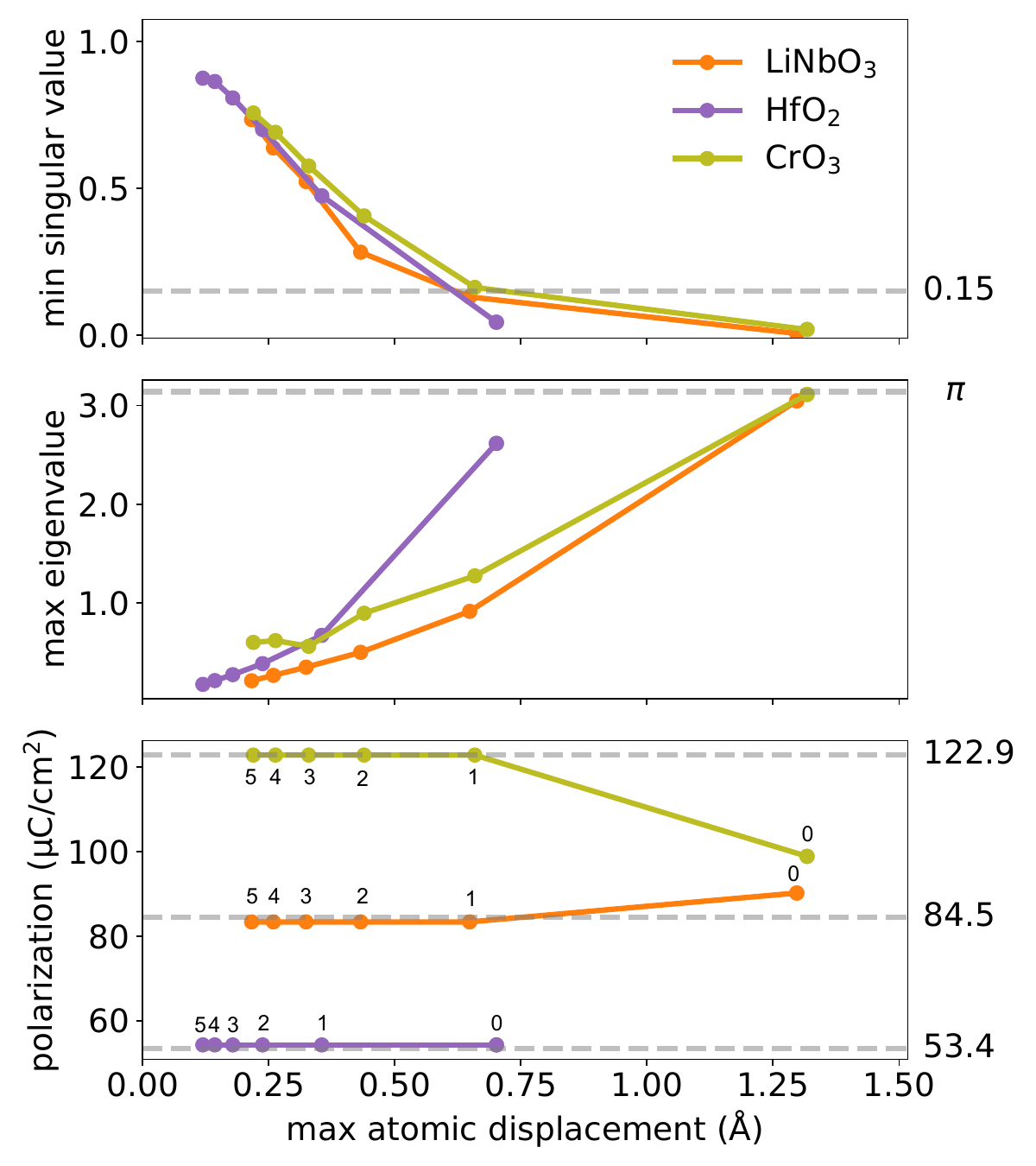}
    \caption{Calculated minimum singular values (top), maximal eigenvalues (middle) and effective polarization (bottom) as functions of the maximum atomic displacement for CrO$_3$, LiNbO$_3$, and HfO$_2$ computed from different numbers of interpolated structures. Results are presented for varying numbers of interpolations, where the points with the largest maximal atomic distance correspond to the polarizations computed directly from the polar and nonpolar reference structures (with the optimal translation applied to the nonpolar structure), and each successive point with a smaller maximal atomic displacement corresponds to one additional interpolated structure. The number of interpolations used in the calculation for each data point is labeled in the bottom panel. For points corresponding to calculations with more than one interpolation, the numbers plotted are the maxima of the maximum atomic displacements across the sets of adjacent interpolated structures used in the intermediate steps of the computation, and likewise for the maxima of the maximum eigenvalues and minima of the minium singular values.
    }
    \label{fig:interpolations}
\end{figure}

Figure \ref{fig:interpolations} demonstrates the impact of using additional interpolated structures on numerical stability for three materials, CrO$_3$, LiNbO$_3$, and HfO$_2$, for which we observe poor numerical stability after applying the translation that minimizes maximal nuclear distances across polar and nonpolar structures. In CrO$_3$, the translation that minimizes the maximum atomic displacement between the polar and nonpolar structures still yields a low minimum singular value (0.020) and a high maximum eigenvalue (3.11) for a displacement of 1.31 Å (Fig. \ref{fig:interpolations}). Introducing four interpolated structures reduces the maximum atomic displacement between adjacent structures to 0.26 \AA, which is below our heuristic value of 0.3 Å. This leads to a significant improvement in numerical stability: the minimum singular value increases to 0.69, and the maximum eigenvalue decreases to 0.62, and we obtain the correct effective polarization value of 122.9 $\mu C / \text{cm}^2$. Similarly, for LiNbO$_3$, we observe that large maximal atomic displacements (1.30 \AA) result in small minimal singular values (0.0063) high maximal eigenvalues (3.05) and an incorrect effective polarization (90.2 $\mu C / \text{cm}^2$). As we add an increasing number of interpolated structures, the numerical stability improves (the minimal singular values increase and the maximal eigenvalues decrease), and the effective polarization value converges to the reference value. In contrast, even under poor numerical conditions (large eigenvalues and small singular values), the effective polarization for HfO$_2$ is computed correctly, indicating that the branch alignment is preserved despite numerical instability. Nevertheless, there are no guarantees that this will always be the case.

These effective polarization values computed from different numbers of interpolated structures demonstrate that while our heuristic—requiring maximum atomic displacement below 0.3 \AA—is more stringent than strictly necessary for some materials, it may be crucial for others. The heuristic provides a conservative guideline for ensuring numerical stability and branch continuity in Berry phase calculations without requiring any DFT input, but our automated workflow will automatically detect and warn users if the maximum eigenvalue is close to $\pi$ or if the minimum singular value is too small ($< 0.15$), regardless of the maximum atomic displacement. Crucially, this heuristic is useful for estimating the number of interpolations required \textit{before} performing any DFT calculations.

\subsection*{Validation of the Berry flux diagonalization approach for computing effective polarization}
\label{subsec:validate_ferroelectrics}

\begin{table}[h!]
  \centering
\begin{tabular}{c 
                  >{\centering\arraybackslash}m{2cm}  
                  >{\centering\arraybackslash}m{2cm} 
                  >{\centering\arraybackslash}m{2cm} 
                  >{\centering\arraybackslash}m{2cm} 
                  } 
    \toprule
    \multicolumn{1}{c}{} & \multicolumn{4}{c}{effective polarization ($\ \mu C / \text{cm}^2$)} \\
    \cmidrule(lr){2-5}
    Material & BFD method (QE) & interpolation method (QE, present work, manual branch tracking) & BFD method (VASP) & 
    interpolation method (VASP, automated branch tracking \cite{smidt2020automatically, ricci2024candidate} )  \\
    \midrule

    BaTiO$_3$ & 45.9 & 45.8 & 45.8 & 45.9 \\
    KNbO$_3$ &  48.0 &  49.3 & 47.3 & 48.7 \\
    PbTiO$_3$ & 115.8  & 118.6 & 116.0 &  116.8 \\
    LiNbO$_3$ & 83.8 &  83.7 & 83.4 & 84.5 \\
    Bi$_2$MoO$_6$ & 76.9 &  80.6 & 75.7 & 7.7* (78.0$^{\dagger}$)\\
    CrO$_3$ & 122.7 & 123.3 & 122.9 & 78.2* (122.9$^{\S}$)  \\
    BiFeO$_3$  & 93.3 & 95.8 &  93.8 & 95.1$^{\ddagger}$\\
    \bottomrule
  \end{tabular}
  \caption{Summary of validation of Berry flux diagonalization (BFD) algorithm with the standard interpolation-based Berry-phase approach computed with \textsc{Quantum ESPRESSO} (columns 2 and 3) and \textsc{VASP} (columns 4 and 5). The interpolation values in the column 5 are reported directly from prior Materials Project (MP)-based  ferroelectric (FE) databases \cite{smidt2020automatically, ricci2024candidate}. We use the same cell parameters for all effective polarization calculations based on the relaxed polar and nonpolar phases as were used in the MP FE database.
  The lattice directions and cell parameters for all materials are reported in Sec. I of the Supplemental Information in Tables I \& II. *Values of polarization reported in Ref. \cite{smidt2020automatically} due to incorrect resolution of the polarization branches. $^{\dagger}$Value of effective polarization calculated with the same method as \cite{smidt2020automatically} but with double the number of interpolated structures to resolve the polarization branch as reported in \cite{ricci2024candidate}.$^{\S}$Value of effective polarization as reported in \cite{smidt2020automatically} by manually resolving the polarization branch.$^{\ddagger}$Value of effective polarization as reported in \cite{neaton2005first}, since Ref. \cite{smidt2020automatically} failed to compute a value of effective polarization due to spurious metallic intermediates. The numbers of interpolations used to obtain these effective polarization values are reported in Table \ref{tab:summary_num_interps}.  Manually resolved polarization branches for the interpolation-based method for \textsc{Quantum ESPRESSO} (column 3) and convergence studies with respect to the number of $k$-points used are reported in Sec. II \& III of the Supplementary Information.}
  \label{tab:summary}
\end{table}

\begin{table}[h!]
  \centering
\begin{tabular}{c 
                  >{\centering\arraybackslash}m{3cm}  
                  >{\centering\arraybackslash}m{3cm} 
                  >{\centering\arraybackslash}m{3cm} 
                  } 
    \toprule
    \multicolumn{1}{c}{} & \multicolumn{3}{c}{\# of interpolations} \\
    \cmidrule(lr){2-4}
    Material & BFD method (QE \& VASP) & interpolation method (QE, present work, manual branch tracking)  & 
   interpolation method (VASP, automated branch tracking \cite{smidt2020automatically, ricci2024candidate})  \\
    \midrule

    BaTiO$_3$ & 0 & 3 & 10 \\
    KNbO$_3$ & 0 & 3 &  10 \\
    PbTiO$_3$ & 1  &  3 & 10   \\
    LiNbO$_3$ & 4  & 3 &   10 \\
    Bi$_2$MoO$_6$ & 2 &  9 & 20  \\
    CrO$_3$ & 4 & 7 &  20 \\
    BiFeO$_3$  & 1 &  5 & 7* \\
    \bottomrule
  \end{tabular}
  \caption{Number of interpolations used to compute the effective polarization values reported in Table \ref{tab:summary}. The number of interpolations required for Berry flux diagonalization (column 2) is calculated according to our heuristics. The number of interpolations for polarization calculations using \textsc{Quantum ESPRESSO} (column 3) is determined by manually tracing branches, where we began with 3 interpolations and increased the number of interpolations as necessary. These manually-resolved branches are reported in Sec. II of the Supplementary Information. The number of interpolations for polarization calculations using VASP is drawn from previous implementations of an automated workflow to compute polarization using the interpolation-based approach \cite{smidt2020automatically, ricci2024candidate}, where the default setting for the number of interpolations is 10, and this number was increased to 20 for CrO$_3$ and Bi$_2$MoO$_6$, since the default 10 interpolations resulted in an incorrect polarization value. This may be viewed as an upper bound on the number of interpolations, instead of a minimum number. *The reference value for polarization for BiFeO$_3$ in the antiferromagnetic configuration using the standard interpolation-based approach is from \cite{neaton2005first}, and we report the number of interpolations that were used in this calculation.}
  \label{tab:summary_num_interps}
\end{table}

We validate the Berry flux diagonalization algorithm by computing effective polarization as the difference in formal polarization between polar and nonpolar structures for a benchmark set of well-known ferroelectrics (BaTiO$_3$, PbTiO$_3$, KNbO$_3$, LiNbO$_3$) as well as a set of proposed polar and known multiferroic materials (Bi$_2$MoO$_6$, CrO$_3$, and BiFeO$_3$) that posed challenges for previous implementations of automated interpolation-based approaches (Table \ref{tab:summary}) \cite{smidt2020automatically, ricci2024candidate}. For each material, wavefunctions are obtained from self-consistent DFT calculations using \textsc{VASP} or \textsc{Quantum ESPRESSO} for polar and nonpolar structures (see Tables 1 \& 2 of the Supplementary Information for cell parameter information). The Berry flux diagonalization method is applied directly to these wavefunctions (with the heuristics explored in the previous sections). For comparison, we compute effective polarization values using the standard interpolation-based approach approach, either manually resolving polarization branches with \textsc{Quantum ESPRESSO} \cite{king1993theory, vanderbilt1998electronic, vanderbilt2000berry} or via the ferroelectric workflow in \textsc{atomate2} \cite{smidt2020automatically, ricci2024candidate} for \textsc{VASP}. 
% Our results are further benchmarked against reference values from \cite{smidt2020automatically, ricci2024candidate, neaton2005first}. 
The effective polarization values reported in Table \ref{tab:summary} are calculated using the PBE functional \cite{perdew1996generalized} with ONCV SG15 pseudopotentials for calculations from wavefunctions using \textsc{Quantum ESPRESSO} and with PAW potentials \cite{kresse1999ultrasoft} for calculations using \textsc{VASP}. Despite these detailed methodological differences, our results obtained using the Berry flux diagonalization algorithm in combination with our real-space heuristics show excellent agreement (with an average percent error of 1.86\% $\pm$ 0.63\% using \textsc{Quantum ESPRESSO} and 1.34\% $\pm$ 0.45\% using \textsc{VASP}) with the interpolation-based reference values, demonstrating that Berry flux diagonalization reliably resolves polarization branches, including in cases where prior automated workflows face significant challenges.

For Bi$_2$MoO$_6$ and CrO$_3$, the standard interpolation-based approach to computing effective polarization has proven sensitive to the number of intermediate structures used; computing the effective polarization in this manner can lead to incorrect results due to unresolved branch structure. In previous work, insufficient interpolation led to substantial errors. For instance, CrO$_3$ was reported to have an effective polarization of $78.2 \ \mu C / \text{cm}^2$ instead of a value of $122.9 \ \mu C / \text{cm}^2$ , the latter of which was reported after manual resolution of the branches \cite{smidt2020automatically}. Similarly, Bi$_2$MoO$_6$ was initially reported to have an effective polarization of $7.7 \ \mu C / \text{cm}^2$ instead of $78.0 \ \mu C / \text{cm}^2$, the latter of which was reported after calculation using the standard approach with denser interpolation scheme \cite{ricci2024candidate}. In contrast, the Berry flux diagonalization method together with our heuristics  enables calculation of effective polarization for these materials using a predetermined (often smaller--see Table \ref{tab:summary_num_interps}) number of interpolations between the polar and nonpolar structures. This greatly reduces the need for multiple interpolated geometries and manual intervention to track branch continuity, while still yielding effective polarization values that match the reference results (See Table \ref{tab:summary}).

BiFeO$_3$ is a multiferroic material that exhibits a G-type antiferromagnetic (AFM) ordering in its ground state structure with spacegroup R3c (161) \cite{neaton2005first}. 
Ref. \cite{smidt2020automatically} did not report the effective polarization of this phase due to erroneous band gap closure and metallic intermediate states for some of interpolated structures. But, it reported a large effective polarization of $124.5 \ \mu C / \text{cm}^2$ for another polar structure of BiFeO$_3$ with spacegroup R3m (167) in the Materials Project database with ferromagnetic ordering, computed using the cubic nonpolar reference phase with spacegroup Pm-3m (221) and the PBE functional. In better agreement with experiment and prior calculations, Ref. \cite{neaton2005first} reported an effective polarization of $95.1  \ \mu C / \text{cm}^2$ for the R3c phase in the AFM ordering, using the standard nonpolar reference with spacegroup R-3c and LDA+U functional to correctly access its semiconducting nature. In this work, we use the same polar and nonpolar phases spacegroups in the same AFM ordering and we find a value that agreed within $2 \ \mu C / \text{cm}^2$ with the value reported in Ref. \cite{neaton2005first} using both \textsc{Quantum ESPRESSO} and \textsc{VASP} (see Table \ref{tab:summary}), with differences attributable to variations in exchange-correlation functional (PBE + U in our work for both codes, versus LSDA + U in \cite{neaton2005first}). (The functional can also impact optimized structural parameters, and to obtain comparisons across codes and materials, we keep the structural parameters, reported in SI Tables 1-2, fixed across calculations.) For variation of the effective polarization with $U$ values for \textsc{Quantum ESPRESSO}, see Sec. IV of the Supplementary Information. For \textsc{VASP}, we use a $U_{\text{eff}} = 5.3 $ eV, which is consistent with the Material Project's recommendation and is the default \textsc{atomate2} \cite{ganose2025atomate2} setting for the BiFeO$_3$.

This validation demonstrates that the Berry flux diagonalization algorithm reliably resolves polarization branches and can avoid failures due to metallic intermediates, making it a robust and efficient alternative to interpolation-based approaches. These advantages are particularly critical for prior automated high-throughput workflows, where 149 out of 413 cases were reported in Ref. \cite{smidt2020automatically} to have failed due to such challenges. 

\subsection*{High-throughput effective polarization calculations with Berry flux diagonalization}
\label{subsec:high_throughput}
We implement a fully automated workflow for Berry flux diagonalization with our real-space heuristics to apply the appropriate translation and number of interpolations using \textsc{VASP} and \textsc{atomate2} \cite{ganose2025atomate2}, and we benchmark the automated Berry flux diagonalization workflow on 176 candidate ferroelectrics from Ref. \cite{ricci2024candidate}. For this comparison, we use the same relaxed polar and nonpolar structures used for the reference effective polarization values and the same functionals (PBE v.54). Across the dataset, agreement with the standard interpolation-based approach is excellent: 154 of the 176 materials differ by less than 1 \% in effective polarization, and 172 of 176 materials differ by less than 10 \%. The root-mean-square difference in effective polarization between the two methods is 0.91 $\sfrac{\mu C}{cm^2}$, demonstrating their overall consistency. Furthermore, our implementation of Berry flux diagonalization requires at most 6 interpolations for any material, with 115 out of 176 materials requiring 1 or fewer and 174 out of 176 requiring 4 or fewer. In contrast, the standard interpolation-based approach used a fixed 10 interpolations by default, and used 20 interpolation in 17 cases when the branch detection algorithm failed with only 10 interpolations. This reduction represents a substantial reduction in computational effort without loss of accuracy. We provide a full table of this high-throughput comparison in Sec. VI of the Supplementary Information.

Only 6 of the 182 candidate ferroelectric materials identified in Ref. \cite{ricci2024candidate} could not be calculated automatically. For CdBiO$_3$, CuReO$_4$, and K$_2$SrCdSb$_2$, we are unable to obtain a large enough band gap for the nonpolar reference structure (using \textsc{atomate2} settings for Brillouin zone integrations, a 0.2 eV Gaussian smearing parameter). Increasing the $k$-point density and using the tetrahedron method, or using a hybrid functional (e.g., HSE), may recover the correct insulating state, and while our workflow flags this situation, future workflows could automatically switch between these strategies when it encounters a gap closure. For LiGaO$_2$, H$_7$ClO$_3$, and RbBe$_2$F$_5$, the automated workflows failed due to parsing issues with the \textsc{VASP} \textsc{WAVECAR} files, which arose due to memory constraints and large file sizes in our calculations. This is due to the fact that our current implementation of Berry flux diagonalization requires wavefunctions over the full Brillouin zone. A Berry flux diagonalization implementation with wavefunctions computed over only the irreducible Brillouin zone, which will result in smaller \textsc{VASP} \textsc{WAVECAR} files would address this issue; we reserve this for future work. Alternatively, one could modify the existing workflow to read in wavefunctions a few $k$-points at a time and compute the overlap matrices for those specified $k$-points independently to circumvent this memory issue.

\begin{figure}
    \centering
    \includegraphics[width=\linewidth]{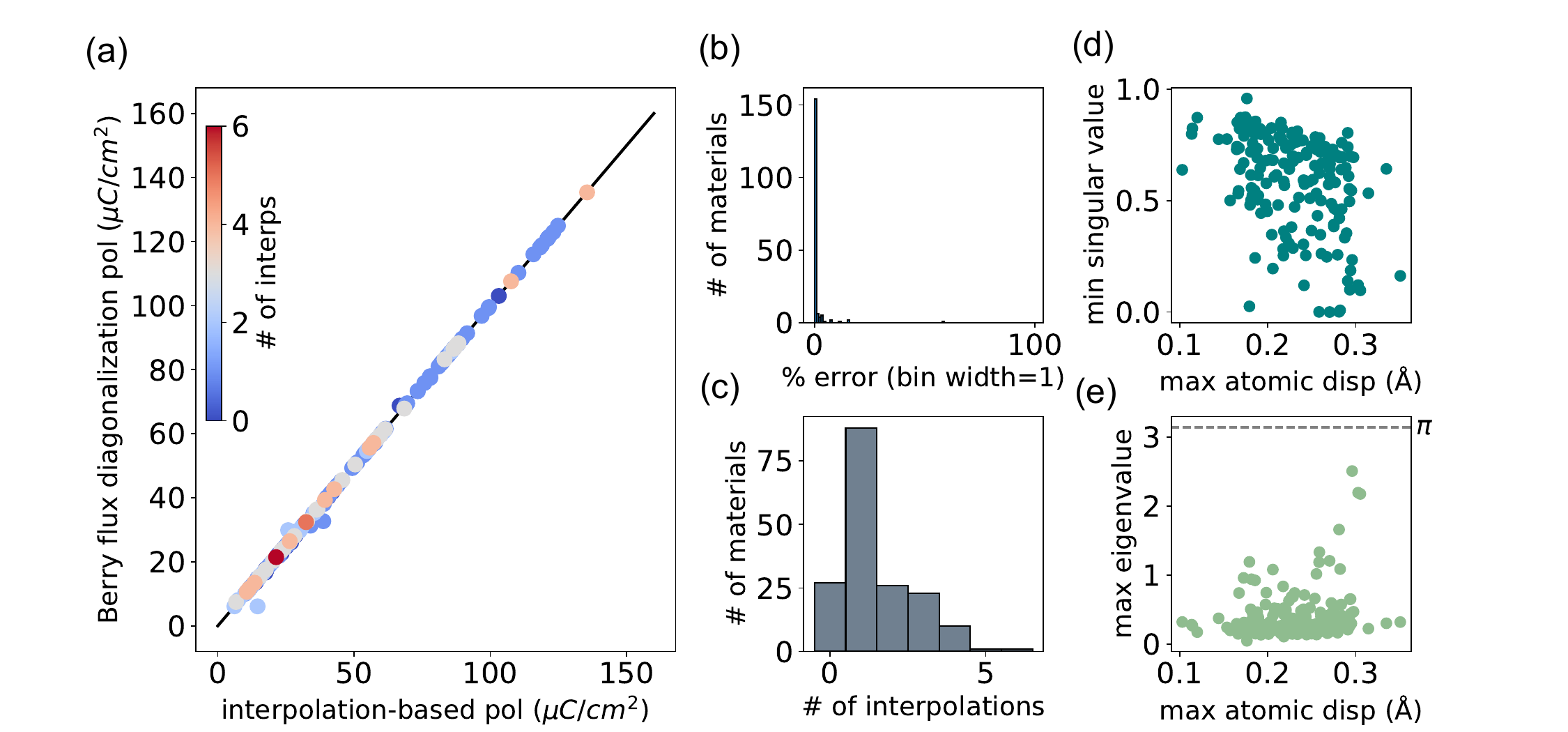}
    \caption{High-throughput calculation results comparing our automated workflow using Berry flux diagonalization with the standard interpolation-based approach for 176 candidate ferroelectric materials. (a) Effective polarization computed with Berry flux diagonalization versus the standard interpolation-based approach. (b) Percent error of the polarization computed with Berry flux digaonalization compared to the standard interpolation-based approach, and (c) the number interpolations used in the Berry flux diagonalization calculation (the standard interpolation-based approach uses 10 or 20 interpolations). (d) Minimum singular values and (e) maximum eigenvalues from the overlap matrices used in the Berry flux diagonalization calculation. When there were multiple interpolation structures, maxima and minima over all structures corresponding to a single material are reported.}
    \label{fig:high-throughput-calc}
\end{figure}

Figure~\ref{fig:high-throughput-calc} summarizes the performance of the automated heuristics for selecting the number of interpolations in effective polarization calculations. Among 176 materials, only four exhibit errors greater than 10~\%: K$_4$Ba$_2$SnBi$_4$ (12.9~\%), Rb$_2$Se$_3$ (15.5~\%), Ba$_2$InSbSe$_5$ (15.8~\%), and Ba$_2$CdAs$_2$ (58.3~\%). Each of these cases is automatically flagged by the workflow due to small minimum singular values ($\leq 0.15$), indicating numerical instability. The user is then prompted to refine the \textit{k}-point mesh, increase the number of interpolations, or validate the results against independent references such as Born effective charges or standard interpolation-based approaches.

For K$_4$Ba$_2$SnBi$_4$, increasing the number of interpolations from 3 to 9 enhances numerical stability (minimum singular value from 0.10 to 0.19) and reduces the effective polarization error to 1.5~\%. In contrast, for Ba$_2$InSbSe$_5$, Rb$_2$Se$_3$, and Ba$_2$CdAs$_2$, increasing the number of interpolations alone does not significantly improve stability or effective polarization accuracy. Instead, a denser \textit{k}-point mesh is required for the self-consistent DFT calculation. The high-throughput workflow employs the default static \textit{k}-point settings from \textsc{atomate2}~\cite{ganose2025atomate2}, which are sufficient for most materials but undersample in these three cases, leading to minimum singular values of the overlap matrices less than the heuristic value of 0.15.

Doubling the \textit{k}-point density in each direction improves the minimum singular values from 0.12 to 0.19 for Ba$_2$InSbSe$_5$, from $6.2\times10^{-4}$ to 0.40 for Rb$_2$Se$_3$, and from $5.1\times10^{-6}$ to 0.34 for Ba$_2$CdAs$_2$, accompanied by improved effective polarization accuracy---percent errors decrease to 8.1~\%, 0.51~\%, and 31.1~\%, respectively, indicating more stringent covergence criteria beyond the default settings can explain the discrepancy. Figure~\ref{fig:high-throughput-calc} reports the results obtained using the automated heuristics and default \textit{k}-point meshes, while both the original and recalculated values are listed in the high-throughput results table in Sec. VI of the Supplementary Information.

These outliers highlight the value of automated flagging based on the intermediate singular values and eigenvalues of the overlap matrices. This capability, combined with an \textit{a priori} estimate of the required number of interpolations based solely on ionic coordinates, makes Berry flux diagonalization both more reliable and more efficient than standard interpolation-based methods, which lack predictive heuristics and automated error detection.

\section*{Discussion}

Berry flux diagonalization represents an advance in the automated computation of effective polarization and can address limitations of interpolation-based approaches in both computational efficiency and reliability. Previous automated workflows for effective polarization calculations rely on interpolation schemes that assume a fixed number of interpolated structures between polar and nonpolar states, regardless of material-specific characteristics \cite{smidt2020automatically, ricci2024candidate}. This framework leads to inefficiencies: in cases where fewer interpolated structures suffice to resolve the polarization path, redundant \textit{ab initio} calculations are performed. Conversely, for materials requiring finer resolution, insufficient interpolation can result in branch misalignment and erroneous effective polarization values. The Berry flux diagonalization method, combined with a set of heuristics based on the ionic coordinates of the polar and nonpolar reference structures presented in this work, overcomes these issues. These heuristics determine the minimal number of interpolations necessary, ensuring both efficiency and accuracy. By monitoring the singular values of the overlap matrices and the eigenvalues of the unitary evolution matrix at each step, the algorithm identifies instances of numerical instability or a failure to enforce gauge alignment, indicating a failure to resolve the polarization branch correctly. In such cases, our workflow halts the calculation and alerts the user, avoiding the propagation of incorrect results. This advance addresses a key challenge in traditional interpolation-based workflows, where branch-jumping behavior is difficult to detect automatically. By explicitly verifying gauge alignment throughout the calculation, this method can provide robust and trustworthy effective polarization values across a wide range of materials.

However, challenges remain for cases where a translation cannot sufficiently minimize the maximal atomic displacement between polar and nonpolar structures. In such cases, additional linear interpolations may fail, similar to interpolation-based Berry phase methods, particularly if the interpolated structures exhibit band gap closures. These limitations highlight the importance of exploring complementary strategies to further enhance the robustness of effective polarization calculations.

Our results indicate that while Berry flux diagonalization can accurately compute polarization for a broad class of materials at a comparable accuracy to the interpolation-based approach, there exists a small class of materials for which the method does not yield stable polarization values, even when the number of interpolations is increased. Crucially, these cases can be automatically identified through monitoring of maximum eigenvalues and minimum singular values, allowing unreliable results to be flagged and excluded. Built-in stability checks ensure that it does not return spurious effective polarization values, making the method reliable and efficient for large-scale screening. 

Looking ahead, several improvements can extend the utility of the Berry flux diagonalization method. Leveraging the symmetry of wavefunctions at different $k$-points could reduce the memory requirements associated with saving wavefunctions across the $k$-mesh for each structure, streamlining high-throughput applications. Additionally, expanding the method's compatibility with other plane-wave-based density functional theory (DFT) codes through the development of parsers for formats beyond \textsc{Quantum ESPRESSO} and \textsc{VASP} will broaden its applicability and adoption. These ongoing developments position the Berry flux diagonalization algorithm as a powerful and versatile tool for advancing ferroelectric materials discovery.

\section*{Methods}

\subsection*{Computational Details}
\label{subsec:computational_details}

The Berry flux diagonalization method employed in this work is an open-source postprocessing algorithm designed for analyzing wavefunctions in a plane-wave basis generated by first-principles density functional theory (DFT) codes. All \textit{ab initio} calculations for generating wavefunctions were performed using \textsc{Quantum ESPRESSO} v7.3 \cite{giannozzi2017advanced} or \textsc{VASP} v6.4.2 with cell inputs obtained from the Materials Project database \cite{jain2013commentary}. For \textsc{Quantum ESPRESSO}, we used SG15 ONCV pseudopotentials and the Perdew-Burke-Ernzerhof (PBE) exchange-correlation functional. Unless otherwise noted, calculations using \textsc{Quantum ESPRESSO} employed a plane-wave kinetic energy cutoff of 100 Ry and a $7 \times 7 \times 7$ $k$-point grid for self-consistent field (SCF) calculations. For non-self-consistent field (NSCF) calculations, we used a denser grid with 21 $k$-points along the $g$-direction and at least 10 $k$-points per string to compute reference polarizations using the ``lberry" tag. For DFT+$U$ calculations on BiFeO$_3$, a Hubbard parameter $U = 3$ eV was applied. For \textsc{VASP}, we used the PBE functional (v54) and the projected augmented wave (PAW) pseudopotentials. The standard interpolation-based approach is implemented in the workflow described in \cite{smidt2020automatically} and in the \textsc{atomate2} code \cite{ganose2025atomate2}. We refer to those papers for more details on the standard interpolation-based workflow, and we use the default DFT parameters for $k$-points and energy cutoffs from the Materials Project for the \textsc{VASP} calculations presented here.

Our implementation of the Berry flux diagonalization method is independent of the specific DFT code used to generate wavefunctions, relying only on the plane-wave format. The algorithm’s implementation is agnostic to internal data structures of the codes and depends solely on \textsc{pymatgen} \cite{ong2013python} for parsing and symmetry operations. In this work, the wavefunction parser for \textsc{Quantum ESPRESSO} was adapted from \textsc{qeschema} \cite{qeschema}, and similar parsers can be developed for other plane-wave-based DFT codes to interface with our framework seamlessly. The wavefunction parser for \textsc{VASP} uses \textsc{pawpyseed} to read PAW DFT wavefunctions and recover the full wavefunctions from the pseudowavefunctions based on their projector functions \cite{bystrom2019pawpyseed}.

\section*{Data and code availability}

All data to support this study and code to compute effective polarization using Berry flux diagonalization will be published under an open-source license and hosted at: \href{https://github.com/pabigail/berry-flux-diag}{https://github.com/pabigail/berry-flux-diag}.

\section*{Acknowledgements}
We thank John Bonini, Stephen Gant, Yongjin Shin, and Vrindaa Somjit for helpful discussions.
This work was funded by the U.S. Department of Energy, Office of Science,
Office of Basic Energy Sciences, Materials Sciences and Engineering Division under
Contract No. DE-AC02-05-CH11231 (Materials Project program KC23MP) for the
development of functional materials. This research used resources of the National Energy Research Scientific Computing Center (NERSC), a U.S. Department of Energy Office of Science User Facility located at Lawrence Berkeley National Laboratory, operated under Contract No. DE-AC02-05-CH11231 using NERSC award BES-matgen and NERSC award DDR-ERCAP0032257.
A.N.P. acknowledges support from the DOE CSGF under Award No. DE-SC0022158.
F.R. acknowledges support from the BEWARE scheme of the Wallonia-Brussels Federation for funding under the European Commission’s Marie Curie-Skłodowska Action (COFUND 847587).

\section*{Author contributions}
A.N.P, F.R., and J.B.N. conceived the study. A.N.P implemented the workflow, performed the calculations, and analyzed the data with input from F.R. and J.B.N. A.N.P drafted the manuscript, and F.R. and J.B.N. worked with A.N.P. to revise and finalize the manuscript.

\section*{Competing interests}
The authors declare no competing interests.

\bibliography{ref}

% --- Include Supplementary Information PDF ---
\clearpage
\includepdf[pages=-]{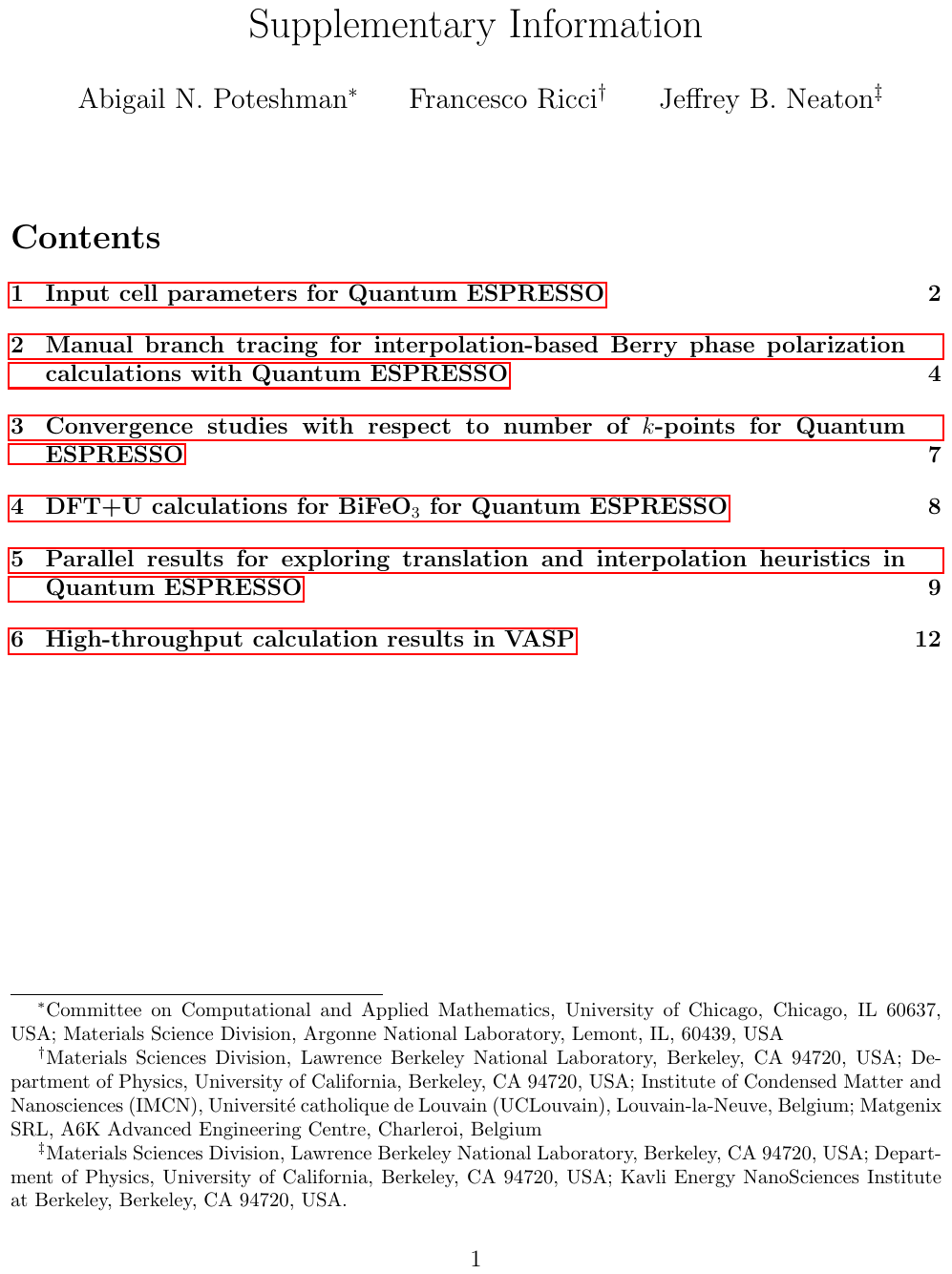}

\end{document}